# Making QCD Lattice Data Accessible and Organized through Advanced Web Interfaces

**Massimo Di Pierro**[*]
*School of Computing - DePaul University - Chicago*
E-mail: mdipierro@cs.depaul.edu

**James Hetrick**
*University of the Pacific*
E-mail: jhetrick@pacific.edu

**Shreyas Cholia**
*Lawrence Berkely Laboratory*
E-mail: scholia@lbl.gov

**David Skinner**
*Lawrence Berkely Laboratory*
E-mail: deskinner@lbl.gov

The Gauge Connection at qcd.nersc.gov is one of the most popular repositories of QCD lattice ensembles. It is used to access 16TB of archived QCD data from the High Performance Storage System (HPSS) at the National Energy Research Scientific Computing Center (NERSC). Here, we present a new web interface for qcd.nersc.gov which allows physicists to browse and search the data, as well as download individual files or entire ensembles in batch. Our system distinguishes itself from others because of its ease of use and web based workflow.

---

[*]Speaker.

## 1. Introduction

The Gauge Connection Lattice QCD data archive has been operated by the National Energy Research Scientific Computing [6] (NERSC) center since 1998. With over 16TBytes of data, the NERSC archive is one of the largest public repositories of lattice gauge ensembles. Its popularity has largely been due to the ease-of-use of its web interface compared to more complex grid based tools.

We are now updating the Gauge Connection to provide a number of modern features, while maintaining the simplicity and ease-of-use of the original. We are adding the ability to search data using tags. We are also providing capabilities to move data easily between the archive and the user's computer, or between two remote computers using both web and grid tools. The archive is now database driven and its pages are dynamically generated in order to facilitate access to the most recent local and remote data. It provides some visulization of the data. New data can easily be uploaded to the archive where it is automatically discovered and cataloged.

The archive can now deliver data in a variety of file formats (including ILDG [5], single and double precision, and FermiQCD [3]) by translating data on the fly after download. We have designed the new website to provide additional capabilities beyond data access, including wiki capabilities to annotate the data and link external work derived the from archive data (derived data, software, tutorials, and publications).

Each gauge ensemble is also associated with a discussion forum allowing registered users to add their own comments. The archive has a more sophisticated access control mechanism and users can have four possible roles: administrators (can manage every aspect), editors (can edit the wiki pages linked to the data), registered users (can download data and comment) and anonymous visitors (can browse the wiki pages and query the data by tags). User management is greatly simplified through the use of a single-sign-on service that allows the user to log in with an existing identity from an external OpenID provider like Google or Yahoo.

## 2. Backend Architecture

The NERSC data is stored on the High Performance Storage System (HPSS), a hierarchical tape storage system designed to manage and access multiple petabytes of data at high speeds [?]. HPSS stores 16 terabytes of Lattice QCD data, mostly generated by the MILC [4] Collaboration. HPSS exposes different data access interfaces including a custom command line client (HSI), a C API, an FTP interface, a parallel FTP interface, as well as a GridFTP interface that can be used via Globus Online. We leverage the HPSS FTP interface to provide web access to the Lattice QCD data.

The system performs nightly introspection of the folder structure (through the HSI and FTP interfaces), reproduces this folder structure in the database, automatically tags data by parsing the file names, groups files with similar name patterns, and publishes the data online. For each folder in HPSS a corresponding web page is generated dynamically. The web structure has the same hierarchy as the folder structure on the HPSS backend.

The new web interface is designed to mimic the iPhone interface and works on both regular browsers and mobile devices.

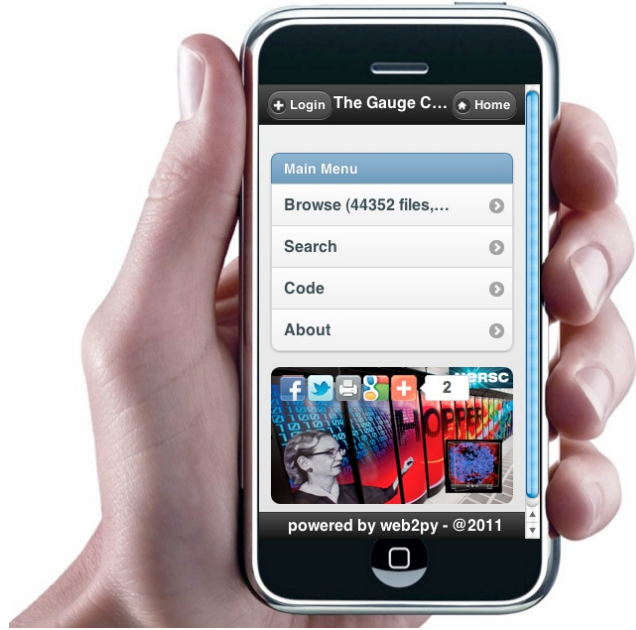

**Figure 1:** The main web page for qcd.nersc.gov allows browsing and searching for gauge ensembles.

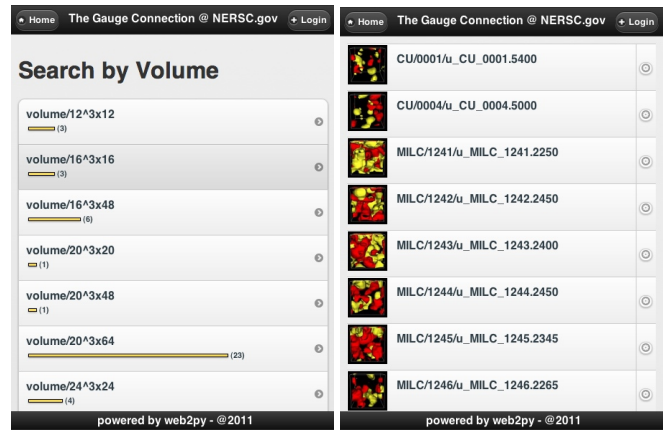

**Figure 2:** The left image show statics by tags. The right image shows topoligical change images from representative elements of various selected ensembles.

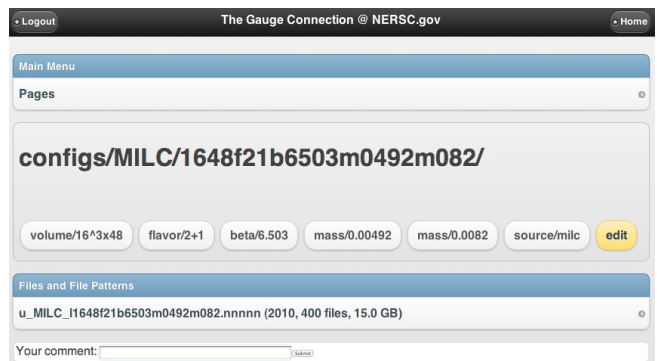

**Figure 3:** The image shows the page corresponding to a folder. The name of the folder is on top, followed by a folder description (optional) and by the tags applied to the folder. Files in the folder are grouped by their prefix (pattern). Patterns are shown at the bottom, above users' comments.

Every page on the QCD site is editable using a wiki syntax similar to wikipedia and registered users can comment on the pages. Both the wiki and comments allow Latex syntax for formulas. Gauge ensembles can be searched by direct browsing of the folder structure or by tag (contributing collaboration, lattice size, beta value, dynamical quark masses, etc.). The system generates interactive charts with statistics by tags.

Some ensembles have been processed off-line to generate meta-data such as a topological charge densities and have been linked to images and movies of said topological charge density. The system is data agnostic and can, in principle, store other data besides gauge ensembles. For example, it can store quark propagators and eigenvalues, examples of which are already in the system. The system uses standard third party web analytics tools to track usage and to geo-tag visitors on a map.

Here is how a simple file access workflow works. The user logs in to the Gauge Connection website (fig. 1) and searches for a desired dataset (fig. 2). When the user finds the ensemble that he/she wants (fig. 3), the user click on the appropriate file to download it. The system checks that the user is logged in, and then pulls the data from tape through the HPSS FTP server. Once the data is available (the tape has been mounted in HPSS), it is streamed back directly to the client browser. Alternatively the user can copy the link to the page for the entire ensemble and pass it to a command line script (provided) that will download every file in the ensemble, one by one, in batch, converting on the fly to the desired format.

## 3. Frontend Architecture

The system is based on web2py [2], a framework for rapid development of secure database driven web applications. It is written in Python and supports many standard databases including Sqlite, MySQL, PostgreSQL, Oracle, MSSQL, Informix, DB2, Sybase, Firebird, and Google Bit Table using a Databae Abstracion Layer (DAL). The DAL generates SQL dynamically and as needed.

The system uses jQuery Mobile for page layout and a custom JavaScript library for charting. It uses the Google Chart API for Latex rendering and the Janrain web service for Single Sign On.

The visualizations of topological charge density are produced offline using Visit (LLNL) and the FermiQCD Visualization Toolkit.

The system has a modular Model-View-Controller design which separates the data representation from the data presentation and from the application workflow. This makes the code compact and easy to maintain. It includes a web based IDE, a web based database management tool and internationalization capabilities.

The complete model for the system consists of just a few lines of code used to describe the data that is stored:

```python
# represents a folder in HPSS
db.define_table('catalog_folder',
    Field('root_id','reference catalog_folder'),
    Field('path'),
    Field('title'),
    Field('header','text'),
    Field('footer','text'),
    Field('pattern_ignore'),
    Field('pattern_group'),
    Field('comments','boolean',default=True))

# a tag to be applied to a folder
db.define_table('tag',
    Field('name'),
    Field('root_id','reference catalog_folder'))

# a file contained in a folder
db.define_table('catalog_file',
    Field('root_id','reference catalog_folder'),
    Field('filename'),
    Field('md5'),
    Field('pattern'),
    Field('extension'),
    Field('size','decimal(20,0)'),
    Field('mtime','datetime'))
```

Each `Field` has a name and a type.

A dispatcher function maps web pages into function calls. These functions are defined in the controller. For example, all web pages of the form

```
http://.../root/<path>
```

Correspond to a `<path>` in the file system and they are mapped into the following function:

```python
@cache(request.env.path_info,60)
def root():
    path = '/'.join(request.args)
    (d,t,f) = (db.catalog_folder, db.tag, db.catalog_file)
    query = (d.path==path) if path else (d.root_id==0)
    page = db(query).select().first() or redirect(URL('error'))
    tags = db(t.root_id==page.id).select()
    subfolders = db(d.root_id==page.id).select()
    patterns = db(f.root_id==page.id).select(
        f.pattern,f.id.count(),f.size.sum(),
        f.mtime.year(), groupby=f.pattern,orderby=f.pattern)
    return locals()
```

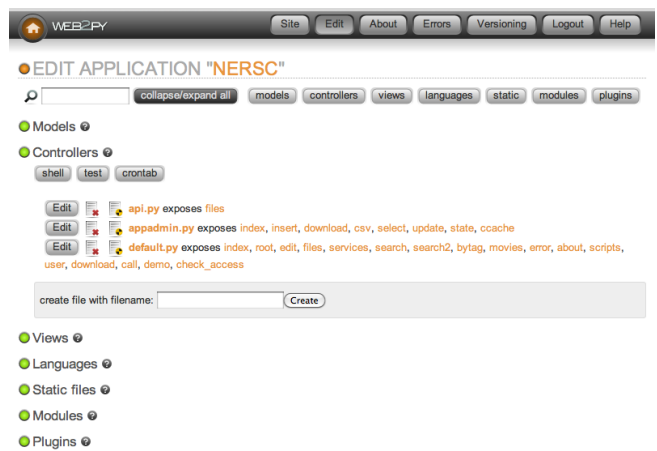

**Figure 4:** The system administrative interface.

This code parses the URL in the request for the path, defines some local variables (d,t,f) referencing the tables, builds a query to fetch the content of the folder: the path, the tags, the subfolders and the patterns (i.e. groups of files with the same prefix). At statements follow the DAL syntax:

```
1 variable = db(query).select(what,how)
```

The @cache(...) decorator instructs the framework to cache each page for 60 seconds. This means the if a page is requested more the once every 60 seconds, it will not hit the database more then one time. This results in better performance.

The system enforces different types of access control. It provides a web based administrative interface only accessible to managers (fig. 4. All downloads require user login.

The system itself is not domain specific and has no knowledge about QCD files and conventions. The domain specific knowledge is in a separate script that runs in background and populates the database from the file/folder structure. This means the system can be used to publish data of a different nature with minimal work. The system retains the original ability to download individual files using the web interface. Since these files can be big, the download requires login in order to prevent Denial of Service attacks.

The system exposes RESTful web services APIs based on the JSON protocol which can be used to access the data programmatically. Some gauge ensembles are comprised of thousands of files. To download them in batch we provide a script qcdutils_get.py, written in Python with no dependencies, available from [1].

In this example we use the script to dowload all files in the *demo* ensemble:

```
1 $ qcdutils_get.py http://qcd.nersc.gov/nersc/api/files/demo
2 http://qcd.nersc.org/nersc/api/files/demo
3 target folder: demo
4 total files to download: 1
5 downloading demo.nersc
```

```
6 demo.nersc 100% |###############| Time: 00:00:00  654.52 K/s
7 completed download: 1/1
```

We now ask the script to convert all data from NERSC to ILDG(32bits):

```
1 $ qcdutils_get.py --convert ildg --float demo/demo.nersc
2 converting: demo/demo.nersc -> demo/demo.nersc.ildg
3   (precision: f, size: 4x8x8x8)
4 100% |###################################|
```

`qcdutils` keeps an internal log of all the operations completed to avoid duplication of work. It remebers what was downloaded/converted and where things are. We can ask `qcdutils` for a log of the completed tasks:

```
1 $ qcdutils_get.py demo/qcdutils.catalog.db
2 demo.nersc created on 2011-06-17T13:42:30.876812
3     [14e7cf9106bfcc16388aeac285ccdad9]
4 demo.nersc.ildg created on 2011-06-17T13:42:34.424604
5     [5a1ae13ddd5cab7ddfe1b17454822ff5]
```

## 4. Conclusions and Outlook

The system allows the user to register any URL and dynamically generates buttons that, when pressed, pass a link to the data at the associated URL. This allows for the creation of third party web services that can feed data directly from the new NERSC web interface allowing for decentralization of services. We can provide tools to help create such services. For example, we can build workflows that interface the data archive with a batch queue on a large computational system.

We envision a future when the different research groups will provide their computing capabilities and their lattice QCD algorithms as web services for the consumption of other members in the collaboration. The NERSC site provides more than just data for these collaborations - it also offers an infrastructure to register those third party services in a transparent way.

**Acknowledgements**

This work was funded by Department of Energy grant DEFC02-06ER41441 and by National Science Foundation grant 0970137.

## References

[1] Massimo Di Pierro. qcdutils, 2010.

[2] Massimo Di Pierro. web2py website, 2010.

[3] Massimo Di Pierro and Jonthan M. Flynn. Lattice QFT with FermiQCD. *PoS*, LAT2005:104, 2006.

[4] S. Gottlieb. Benchmarking and tuning the MILC code on clusters and supercomputers. *Nuclear Physics B Proceedings Supplements*, 106:1031–1033, 2001.

[5] C. Maynard. International Lattice Data Grid: Turn on, plug in,and download. In *Symposium on Lattice Field Theory*, 2009.

[6] U.S. DOE. Nersc website, 2010.